\documentclass[twocolumn]{aastex63}
\usepackage{tabularx}
\usepackage{multirow}
\usepackage{longtable}
\usepackage{dcolumn}
\usepackage{lineno}
\usepackage{amsmath}
\usepackage{mathrsfs}

\usepackage{cancel}

  \newcommand{\mytextdeleted}[1]{}

\newcommand\etal{\it{et al.} \rm }

\submitjournal{ApJS}

\shorttitle{Spin Parity VI}
\shortauthors{Iye \etal}

\begin{document}

\title{Spin Parity of Spiral Galaxies VI -- A Search for Dynamical Memory in the
Spin Distribution of Galaxies in HSC WIDE Survey Regions}

\correspondingauthor{Masanori Iye}
\author[0000-0002-5634-7770]{Masanori Iye}
\email{masanori.iye2@gmail.com}
\affil{National Astronomical Observatory of Japan, Osawa 2-21-1, Mitaka, Tokyo
181-8588 Japan}
\affil{The Japan Academy,7-32 Ueno Park, Taito-ku, Tokyo 110--0007, Japan}

\author[0000-0001-7550-2281]{Masafumi Yagi}
\email{yagi.masafumi@nao.ac.jp}
\affil{National Astronomical Observatory of Japan, Osawa 2-21-1, Mitaka, Tokyo
181-8588 Japan}

\begin{abstract}
We analyzed the distribution of spin parity in spiral galaxies using the HSC DR2 data. The spiral winding parity of disk galaxies, observed as S-spiral or Z-spiral projected onto the sky plane, provides robust information on the sign of the line-of-sight component of their spin vectors, specifically whether the spin vector points toward or away from us.
The distribution of 49,494 S/Z annotated spirals with spectroscopic redshift (0.05 $\le z$) was analyzed for 46,247 fiducial cubic search volumes of various sizes, 20--200 Mpc, deployed in the 3D supergalactic
coordinates. We counted the number of S-spirals and Z-spirals in each cube, evaluated the binomial probability of the observed S/Z imbalance, and identified statistically anomalous cube candidates.

The observed cumulative distribution functions for the 256 sets of cubes are in good agreement with the theoretical binomial distribution and with those obtained from 1000 Monte Carlo realizations assuming random S/Z spin assignments.  The number of statistically anomalous cubes is also comparable to that expected from the random assignments. These results indicate that the spin-vector distribution of spiral galaxies is consistent with statistical randomness expected from the standard cosmological model of structure formation.

\end{abstract}


\section{Introduction} \label{sec:introduction}

Whether the spin vector distribution of galaxies in the substructures of the universe is random has been a subject of interest in scenarios related to structure and galaxy formation \citep{Peebles1969, Zeldovich1970, White1984}. 
In the standard cosmological model, large-scale structure arises from the gravitational amplification of primordial density perturbations. Although vorticity decays with cosmic expansion at linear order \citep{Bardeen1980}, it can be generated at higher orders \citep{Christopherson2011}. Galaxy angular momentum is commonly attributed to tidal torques; however, a definitive theoretical and observational validation has not yet been achieved.

A popular observational approach to study the spin distribution of galaxies is to use the ratio of the minor to major axes of galaxies, along with the position angle of the major axis, to study the distribution of the tangential component of the spin vector of galaxies.
\citet{Pajowska2019} studied 247 Abell clusters of galaxies and observed that their orientations were not random. \citet{Trujillo2006} showed that the spin vectors of galaxies in the Sloan Digital Sky Survey (SDSS; \citealp{York2000})
and the Two Degree Field Galaxy Redshift Survey (2dFGRS; \citet{Colless2001})
tend to be perpendicular to the radial direction pointing to the void.
\citet{Varela2012} studied SDSS DR7 galaxies and observed that galaxies around cosmic voids tend to have spin vectors aligned radially to the voids.
\citet{Zhang2013} studied SDSS DR7 galaxies and observed a statistical alignment
of the galaxy axis with the filament axis or sheet plane.
\citet{Abdullah2014} reported that there was no directional correlation between galaxy spin-axes and filamentary structure in the distribution of galaxies at $0.0046 \leq z \leq 0.0073$ they studied. \mytextdeleted{, whereas \citet{Mesa2018} studied 25,965 SDSS galaxy pairs at $z \le 0.1$ with separation less than 100$h^{-1}$ kpc and observed that the axis connecting the pairs is statistically in alignment with the axis of the background filament structure.}
\citet{Lopez2021} observed that whereas the major axis undergoes a considerable change, from being perpendicular to the filament axis initially to being aligned at present, the spin orientation exhibits a mild evolution, becoming preferentially perpendicular to the filament axis. 
\citet{Motloch2022} observed that galaxies tend to be preferentially oriented
face-on when the major axis of the initial tidal field is aligned with the line
of sight.

In the context of cosmological galaxy–shape correlations, the so-called intrinsic alignments (IAs), galaxy spin is modeled as being driven by the large-scale tidal field through tidal torque effects \citep{Hirata2004}.
Observational efforts to detect or constrain such IA signals have made use of the ellipticity of star-forming disk galaxies.
In practice, these analyses have typically relied on samples of blue galaxies, which broadly correspond to the spiral population.
\citet{Hirata2007} investigated the intrinsic alignment of blue galaxies using SDSS imaging data and reported no significant detection.
This result was revisited by \citet{Mandelbaum2011}, who combined SDSS shape measurements with spectroscopy from the WiggleZ survey, again finding no alignment but placing tight upper limits.
\citet{Li2013} observed an alignment signal in the two-point correlation
function, which probes the dependence of galaxy clustering at a given separation in redshift space on the projected angle between the orientation of galaxies and the line connecting them to other galaxies.
More recently, \citet{Johnston2019} analyzed blue galaxies from the KiDS+GAMA survey, complemented with SDSS spectroscopy, and reported a null detection while providing stringent constraints on the alignment amplitude.

While studies of galaxy shape and orientation make use of the distribution of the tangential component of the spin vector, information on the radial component of the spin vector has not been incorporated to date.
Moreover, those analyses inevitably involve non-trivial measurement errors. Calibrating the associated statistical analyses is not
straightforward, and drawing definitive conclusions is often challenging.
Another difficulty of this approach is the fourfold degeneracy in the 3D direction of its spin vector.

Another independent and complementary approach is to use the winding direction of the spiral structure of disk galaxies to study the distribution of the radial component of the spin vector of galaxies. There is a consensus that these structures are trailing spirals, rather than leading spirals. In fact, \citet{Iye2019} affirmed that all 146 spiral galaxies for which the near side of the tilted disk is identifiable have trailing spirals. This presupposition provides a firm basis to allow the judgment of the sign of the line-of-sight component of the spin vector of a spiral galaxy, simply by judging whether the spiral pattern projected on the sky is winding ``S-spiral'' or ``Z-spiral.'' The datum used here is merely a single bit of information for each galaxy. Still, it provides a robust set of observational details.

The Galaxy Zoo project \citep{Lintott2011, Willett2013}, which began in 2007, generated a morphological catalog, GZ1, of 893,212 SDSS galaxies at redshifts up to 0.25 through the visual classification voting of many volunteer citizens. A total of 18,454 S-spirals ($P_{ACW}\geq0.8$) and 17,256 Z-spirals ($P_{\rm CW}\geq0.8$) were identified in catalog GZ1.
\citet{Hayes2017} noted that the number of S-spirals in the catalog is
significantly larger than that of Z-spirals by approximately 10$\%$. As such, asymmetry is complex to explain from a physical viewpoint; the authors suspected that this anisotropy might be due to human recognition bias, which is yet to be clarified. \citet{Patel2024} made reanalyses of publicly available data sets with spin classifications and indicated consistency with isotropy to within 3 $\sigma$.

In this paper, we investigate deviations from random S/Z assignment models across a range of spatial scales, using HSC spiral galaxies with S/Z classifications and spectroscopic redshifts. Section 2 describes the data used in the present paper. We start by deploying search cubes to cover the observed domain in Section 3.1 and calculate the binomial probabilities of finding the observed number imbalance for each region in Section 3.2. We construct a variety of sets of cubes and compare the observed one-sided cumulative distribution function with that of the theoretical binomial distribution model to calculate the Kolmogorov-Smirnov statistic in Section 3.3. We evaluate the level of statistical significance for each set using 1000 Monte-Carlo simulations in Section 3.4.  The bizarrest cube and the bizarrest hemicube are described in Sections 3.5 and 3.6, respectively, and their statistics are examined. To our knowledge, this is the first study to examine the three-dimensional spatial distribution of spiral galaxy spin vectors.

\section{HSC WIDE Sample of Galaxies and Their Spin Annotation}
\label{sec:statistics}

\subsection{Tadaki2020's  Analysis}

We use the data release 2 of the HSC-SSP for the WIDE fields \citep{Aihara2019}, which covered four equatorial regions in spring and autumn (https://hsc.mtk.nao.ac.jp/ssp/survey/\#survey\_fields). The four fields labeled W05H, W12H, W15H, and W23H cover equatorial fields with field centers at around 5h, 12h, 15h, and 23h in the right ascension as shown in Table \ref{tab: SurveyFields}. A total of 561,251 galaxies are identified in these HSC-SSP WIDE fields over $\sim 320 \ deg^{2}$ on the sky.

\begin{table} 
\centering
\caption{HSC WIDE Survey Fields}\vspace{-7pt}
\begin{tabular}{ccccc} \hline
&W09H&W12H&W15H&W23H\\ \hline
R.A.&128.9:154.8&174.0:188.9&206.6:226.0&331.0:3.7\\ \hline
Decl&-1.4:7.3&-1.5:4.5&-1.5:0.5&-1.5:3.0\\ \hline
\end{tabular}
\label{tab: SurveyFields}
\end{table}

Using a convolutional neural network (CNN) program trained by tutorial data,
\citet{Tadaki2020} classified all the galaxies into three classes: 37,917
S-spirals, 38,718 Z-spirals, and 484,616 non-spirals. The classification was
based on the largest $p$-value among the three classes assigned to each galaxy
by the CNN. The training of this CNN algorithm is based on a verification sample
of 54,475 galaxies in the XMM-LSS field, observed with the same instrumental
settings. Visual annotations of clear 1,447 S-spirals and clear 1,382 Z-spirals
were made by two of the authors of this paper (MI and HF). The tutorial data set
was enhanced by adding flipped and rotated images to eliminate potential biases.

\subsection{Variable Threshold of Annotation}
We first reexamine the annotation scheme of \citet{Tadaki2020} by investigating its dependence on the annotation threshold $p$-value. By setting a threshold $p$-value, we re-annotate the galaxy to the three classes (S, Z, N) only if the highest $p$-value among the three classes exceeds this threshold, and put the galaxy into the fourth unclassified category (U) when none of the $p$-values for the three classes exceed the adopted threshold. 
With this scheme, we have 12 groups of samples for the verification sample regarding the matching between the three eye annotation classes (S, Z, N) and the four CNN annotation classes (S, Z, N, U), namely NN, NS, NZ, SN, SS, SZ, ZN, ZS, ZZ, NU, SU, and ZU.

We define the maximum fractional annotation error by
$\epsilon_{\rm annot}=(n_{\rm SN}+n_{\rm ZN}+n_{\rm NS}+n_{\rm NZ}+n_{\rm SZ}+n_{\rm ZS})/n_{\rm total}$. With this definition of error, \citet{Tadaki2020} reported that their sample, with threshold $p=0.358$, which we call $P_{\rm 36}$ sample, had up to 3.99\% annotation error.

Table \ref{tab: p-value dependence} shows the number statistics of the S/Z annotation by CNN for 49,494 galaxies observed in HSC WIDE survey fields for some representative threshold $p$-value samples for the entire spectroscopic redshift sample of our new catalog. 

The annotation error $\epsilon_{\rm annot}$ decreases from 3.99\% for $P_{\rm 36}$ sample to 0.6\% for $P_{\rm 98}$ sample with the threshhold $p$-value 0.98. However, applying a stringent threshold $p$-value to reduce annotation error is accompanied by a decrease in the number of annotated spirals, to 53\% in the above case, increasing Poissonian fractional count error, $\epsilon_{\rm Poiss}=1/\sqrt{n_{\rm all}}$, where $n_{\rm all} = n_{\rm S} + n_{\rm Z}$ is the total number of S-spirals $n_{\rm S}$ and Z-spirals $n_{\rm Z}$. 
We chose, therefore, $P_{\rm 36}$ sample for our further study.

Note that the Poissonian error dominates over the annotation error, $\epsilon_{\rm annot} < \epsilon_{\rm Poiss}$, for galaxy samples with $n_{\rm all} < n_{\rm all}^{\rm crit} = \epsilon_{\rm annot}^{-2}$.

Note also that the annotation error was estimated from the validation sample, while the Poisson error was calculated from the HSC-annotated data.

\subsection{Revised Spin Catalog for HSC WIDE Fields}
Among the 37,917 S-spirals and 38,719 Z-spirals annotated by Tadaki+2020, the spectroscopic redshifts were known only for 15,757 of them in the PDR2 catalog of HSC. 
We cross-matched all 76,636 S/Z-annotated spiral galaxies with the available catalogs, together with our new spectroscopic observations from the Subaru Prime Focus Spectrograph, thereby increasing the number of spirals with spectroscopic redshifts to 49,494. Details of the procedures used to compile the galaxy spin catalog in this study and the machine-readable Revised Spin Catalog of HSC WIDE Fields are provided in Appendix A.

The redshifts of 95.6\% of them were in the range $0.05 < z <0.5$. 
During this process, we identified and cleaned multiple entries in the Tadaki catalog. We checked the DESI online spectrum, where the DESI and SDSS redshift measurements are inconsistent, and determined a relevant redshift value. 
The spectra of galaxies at large redshifts ($z \ge 0.8$) with S/Z annotations were also checked individually, as the S/Z annotation would imply that they are not too distant objects. 
The redshifts of objects at $z < 0.05$, where Galactic objects could well contaminate \citep{Yagi2025}, were also individually checked.

\begin{table}[] 
\centering
\caption{Annotation results for the HSC WIDE spectroscopic sample at various $p$-value thresholds.}
\begin{tabular}{cccc} \hline
&$P_{36}$&$P_{98}$&$P_{999}$\\ \hline
$n_{\rm S}$&24603&14961&7014\\ \hline
$n_{\rm Z}$&24891&14933&7045\\ \hline
$n_{\rm U}$&0&19600&35435\\ \hline
$\epsilon_{\rm annot}$&3.99\%&0.60\%&0.06\%\\ \hline
\end{tabular}
\label{tab: p-value dependence}
\end{table}

\subsection{Application to HSC WIDE fields}

A sample larger than $n_{\rm all}^{\rm crit}$ is required to ensure that the analysis is not limited by annotation error. For the sample $P_{\rm 36}$, the critical number is $n_{\rm all}^{\rm crit} = 628$. Our new catalog is still not large enough to ensure $n_{\rm all} \ge n_{\rm all}^{\rm crit}$ in most of the searched cubes. Therefore, we employ the $P_{36}$ sample for our analysis in this paper.

The distances to galaxies are derived from the observed spectroscopic redshift, assuming Planck18 $\Lambda$CDM cosmology \citep{Aghanim2020} with Hubble constant $H_{0} = 67.4 \rm km\ s^{-1} \ Mpc^{-1}$, matter density $\Omega_{\rm m} = 0.315$, and matter
fluctuation amplitude $\sigma_{8}=0.811$.

We do not take the peculiar velocities of galaxies into account and locate each galaxy in the Cartesian supergalactic coordinates (SGX, SGY, SGZ). In the following, we use these 49,494 spirals for which the 3D positions and S/Z annotations are securely available.

\section{Search Cube Analysis}

\subsection{Search Cube Deployment}

\begin{figure}[h] 
\centering
\includegraphics[scale=0.25]{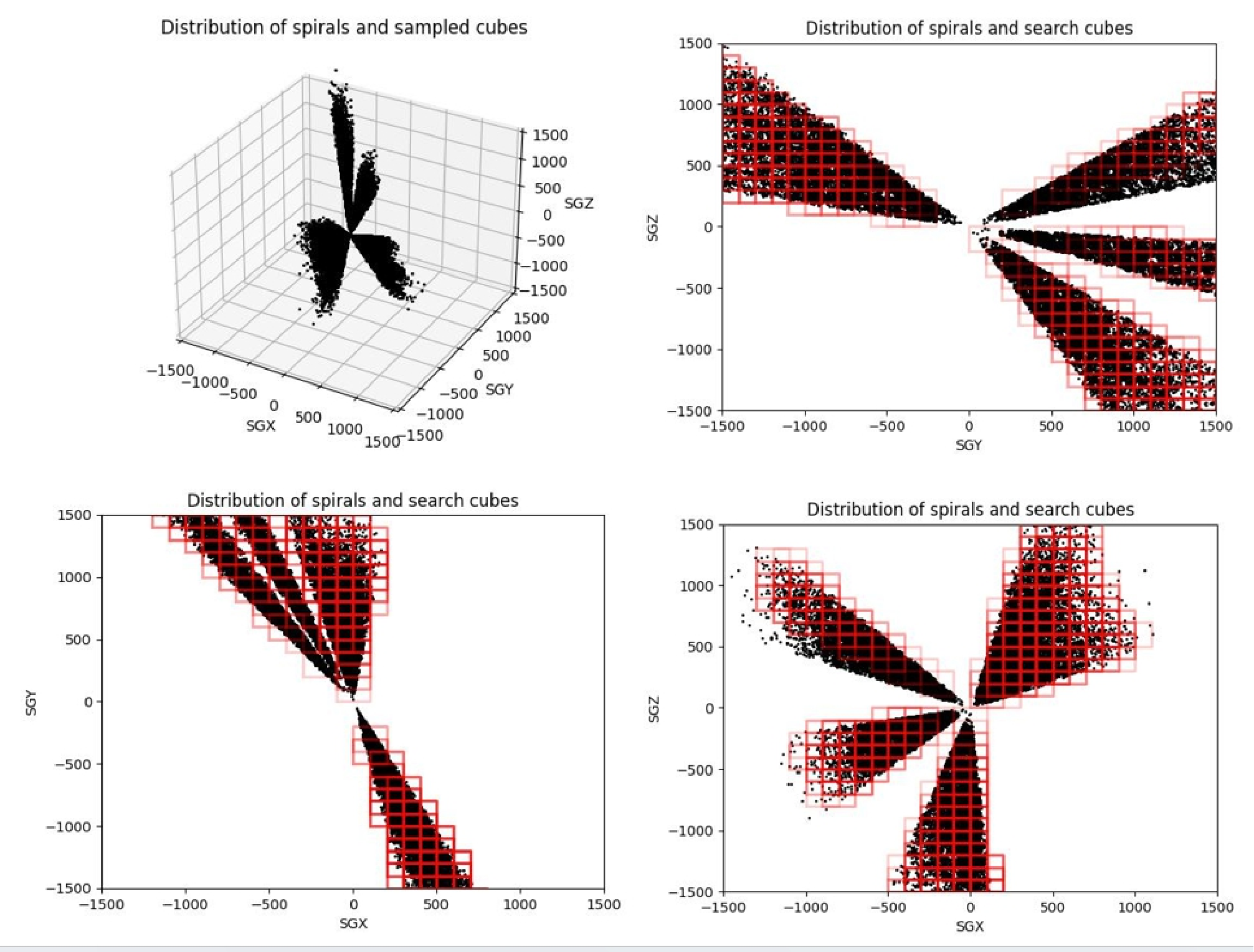}
\caption{Spatial distribution of sampled spiral galaxies (black dots) in supergalactic coordinates, overlaid with search cubes of radius 100 Mpc (red), shown within the inner 3 Gpc cubic domain.}
\label{fig: SearchCubes}
\end{figure}

To investigate potential anisotropies on cosmological scales, we construct three-dimensional arrays of search cubes with uniform size within a 6-Gpc MegaCube that encompasses the entire observed volume of the four HSC WIDE regions. In each 3D array, the centers $(x_{c}, y_{c}, z_{c})$ of adjacent cubes are separated by $2R$, where $R$ corresponds to the radius of a sphere inscribed in a cube with side length $2R$. We generate 10 cube arrays with radii in the range from $10 \le R \le 100$ Mpc to check the dependency on searched size of cube volumes. Hereafter, we use $N$ to denote the number of cubes and $n$ to denote the number of galaxies within a cube.

Further, to mitigate the risk of missing anisotropies located near the boundaries of the search cubes, we construct eight 3D arrays (Array 1 to Array 8) for each radius $R$, shifting the grid by $R$ rather than the full spacing of $2R$. This cube deployment configuration ensures that each sub-cube of size $R$ is sampled up to eight times by neighboring cubes of side length $2R$. The total number of cubes in each Array is in the range 5700- 5835, about 1/8 of the 46247.

The total number of cube sets, denoted $\mathscr N^{\rm T}_{\rm sets}$ with the minimum number threshold $n_{\rm lim}$ = 11 to select cubes, is therefore, $\mathscr N^{\rm T}_{\rm sets} = 80$.

In addition, we generated cube set with three more different values of $n_{\rm lim}$ to see the dependence of $n_{\rm all}$. The total number of cube sets decreases with increasing $n_{\rm lim}$. We omit sets with $N_{\rm cubes} < 11$, and the grand total number of cube sets with $n_{lim} = 11, 30, 100,$ and $300$ is $\mathscr N^{\rm G}_{\rm sets} = 256$. We assigned a tentative cube set ID number $k$ with $k=1, \ldots, 256$.

Note that cubes in an array set do not share the same galaxy. For statistical consistency, each array is analyzed separately to avoid multiple counting of the same galaxies.

Figure \ref{fig: SearchCubes} shows the 3D bird's-eye view and the projected distributions in the super galactic coordinates of studied galaxies (black dots) and the 200 Mpc searched cubes with $n_{\rm all} \ge 11$. Most sub-cubes of 100 Mpc size (thick red) are multiply contained in adjacent cubes. Singly sampled sub-cubes (thin red) are seen at the periphery of the searched fields.

\subsection{Binomial Probabilities of Search Cubes}

In the absence of spatial correlations among spiral galaxy spin axes, and assuming that spectroscopic observations are independent of spin parity, the resulting number counts should follow a binomial distribution with $p=0.5$.
The binomial probability of finding $n^{i}_{\rm S}$ S-spirals among $n^{i}_{\rm all}$ spirals is
\begin{equation}
B(n^{i}_{\rm S},n^{i}_{\rm all},0.5)=\binom{n^{i}_{\rm all}}{n^{i}_{\rm S}}\left(\frac{1}{2}\right)^{n^{i}_{\rm all}}.
\end{equation}
We define the one-sided tail probability $p_{\rm S}$ by
\begin{eqnarray}
p^{i}_{\rm S}
= P(k \le n^{i}_{\rm S}) 
= \sum^{n^{i}_{\rm S}}_{k=0}B(k, n^{i}_{\mathrm{\rm all}}, 0.5),
\end{eqnarray}
\noindent
which is the probability that $n_{\rm S}$ takes a value equal to or smaller than the observed number $n_{\rm S}^{i}$ under the binomial distribution.
Similarly, we characterize each search cube by a two-sided tail probability $p^{i}$ defined by
\begin{eqnarray}
p^{i} 
=
P(k \le n^{i}_{\rm S} \hspace{3mm} \rm or \hspace{3mm} \it k \ge n^{i}_{\rm all}-n^{i}_{\rm S})\nonumber \\
=
2 \sum_{k=0}^{\min(n^{i}_{\rm S},n^{i}_{\rm Z})}B(k, n^{i}_{\rm all}, 0.5), \hspace{5mm} \rm if \hspace{2mm} n^{i}_{\rm S} \ne n^{i}_{\rm Z}.
\end{eqnarray}

Note that if $n^{i}_{\rm S} = n^{i}_{\rm Z} = n^{i}_{\rm all}/2$, Eq. (3) incorrectly counts the $k=n^{i}_{\rm all}/2$ probability twice and $p^{i}$ gets larger than 1. For this case, $p^{i}$ should be set to 1.

Thus, $p^{i}$ is the two-sided tail probability that the cube exhibits a number asymmetry equal to or larger than the observed one. The binomial distribution approaches a normal distribution in the limit $n_{all} \to \infty$. However, the one-sided probability $p_{\rm S}^{i}$ takes values in the range $2^{-n_{\rm all}} \le p_{\rm S}^{i} \le 1$, whereas the two-sided probability $p^{i}$ lies in the range $2^{-n_{\rm all}+1} \le p^{i} \le 1$.


\begin{figure}[h] 
\centering
\includegraphics[scale=0.27]{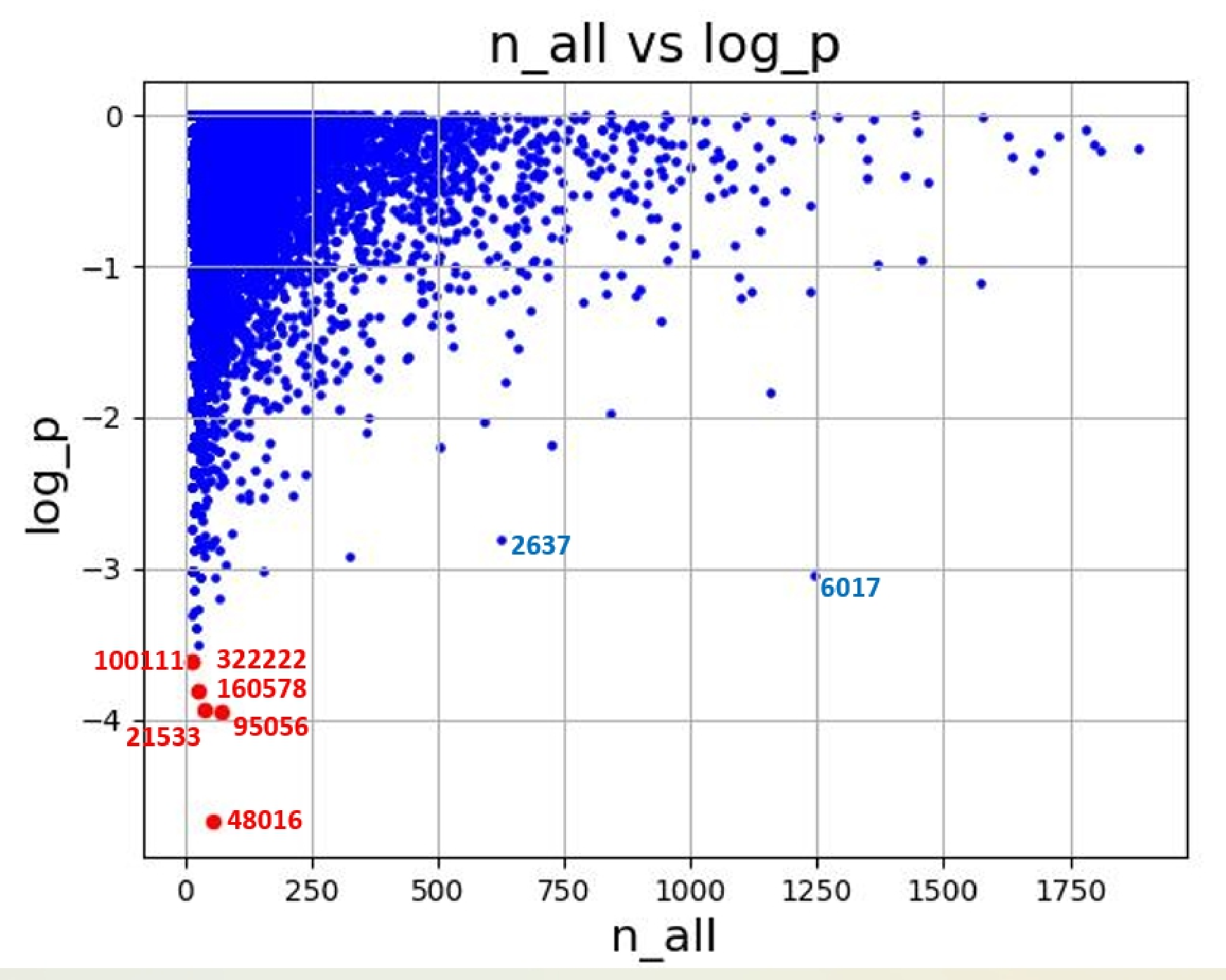}
\caption{Distribution of $p$ as a function of $n_{\rm all}$ for the 46,247 search cubes (blue points). The six ``highly bizarre'' cubes with $b  < 0.1$ (red circles), listed in Table \ref{tab: BizarreCubes}, are labelled.  Cube 2637 and 6017, which contain particularly large values of $n_{\rm all}$, are also highlighted. \\
}
\label{fig: nall-logp}
\end{figure}

Figure \ref{fig: nall-logp} shows the distribution of measured $p^{i}$ \rm as a function of $n^{i}_{\rm all} =n^{i}_{\rm S}+n^{i}_{\rm Z}$ for $11 \leq
n^{i}_{\rm all} \leq 1,887$ for the 46,247 cubes studied.

Table \ref{tab: SearchCubes} shows the number of searched cubes as a function of the counted number of S/Z annotated spirals $n_{\rm all}$ for each $R$ The column with $n_{\rm all} \ge 628$  shows the number of cubes for which $\epsilon_{\rm Poiss} < \epsilon_{\rm annot}$ in our $P_{36}$ sample.

\begin{table}[] 
\centering
\caption{Numbers of search cubes $N_{\rm cubes} (R, n_{\rm lim}$)  of ``radius'' $R$ (listed at left), in which the total number of enclosed S/Z spirals $n_{\rm all} \ge n_{\rm lim}$. The number of independent cubes in each array $N^{Array}_{\rm cubes}$ is roughly 1/8 of $N_{\rm cubes}$.}
\begin{tabular}{cccccc} \hline
$N_{lim}$&$11$&$30$&$100$&$300$&$628$\\ \hline
100&833&613&376&203&103\\ 75&1492&1050&590&250&76\\ \hline
60&2268&1489&736&231&32\\ \hline
50&3092&1942&819&173&3\\ \hline 
37.5&4759&2590&780&45&0\\ \hline 
30&6280&2941&555&10&0\\ \hline
25&7398&2928&323&0&0\\ \hline 
20&8273&2822&118&0&0\\ \hline
15&7703&1277&21&0&0\\ \hline
10&4149&3000&0&0&0\\ \hline 
Total&46,247&17439&4318&912&214\\ \hline 
\end{tabular}
\label{tab: SearchCubes}
\end{table}

\begin{figure}[h] 
\centering
\includegraphics[scale=0.27]{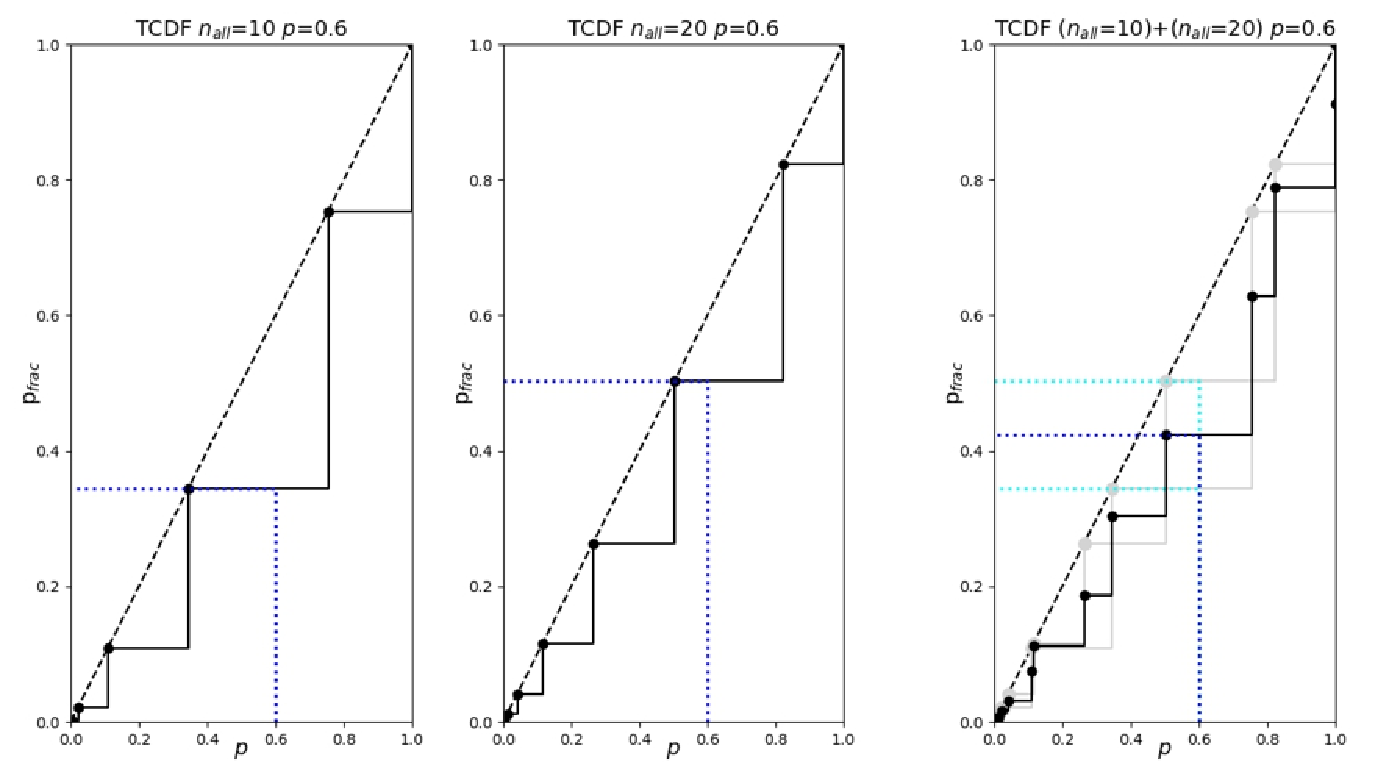}
\caption{
The left and middle panels show the theoretical binomial  $OCDF$ for cubes with $n_{all} = 10$ and $n_{all} = 20$, respectively, where the steps of the distribution intersect the diagonal dashed line corresponding to the normal distribution. The right panel shows the $OCDF$ for a combined set of the two cubes, in which the steps deviate from the normal distribution line. The probability of obtaining an asymmetry with $p_{S} \le 0.3$, for example, is obtained by reading the corresponding vertical value of the $OCDF$ in each panel. Note that the $OCDF$ for the theoretical binomial distribution lies below the normal distribution line due to its discrete nature.}
\label{fig: OCDF}
\end{figure}

\begin{figure}[b] 
\centering
\includegraphics[scale=0.38]{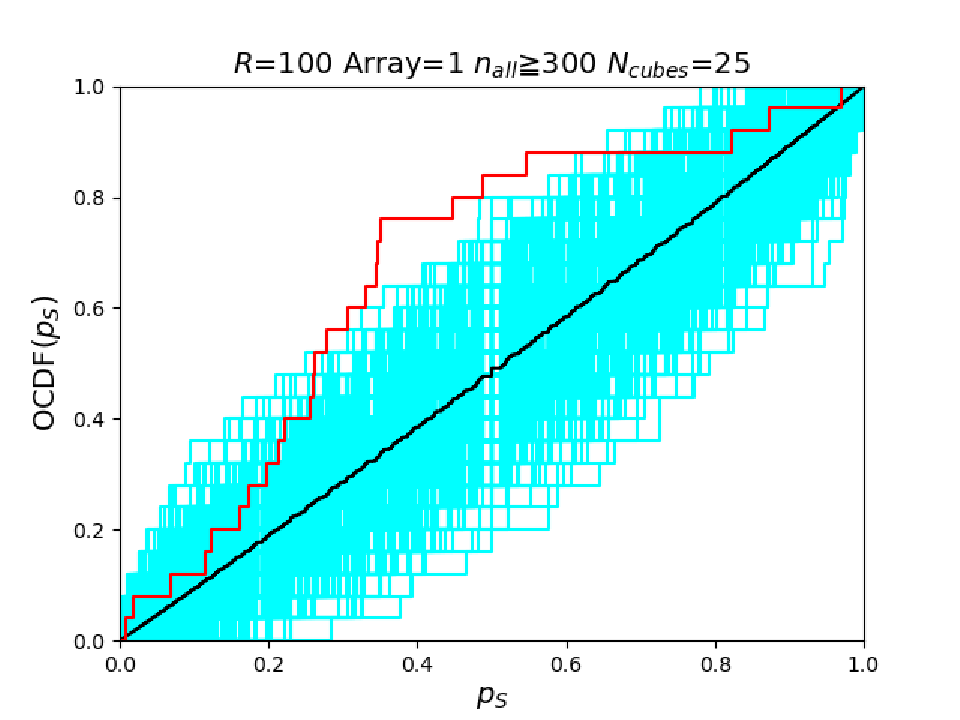}
\includegraphics[scale=0.38]{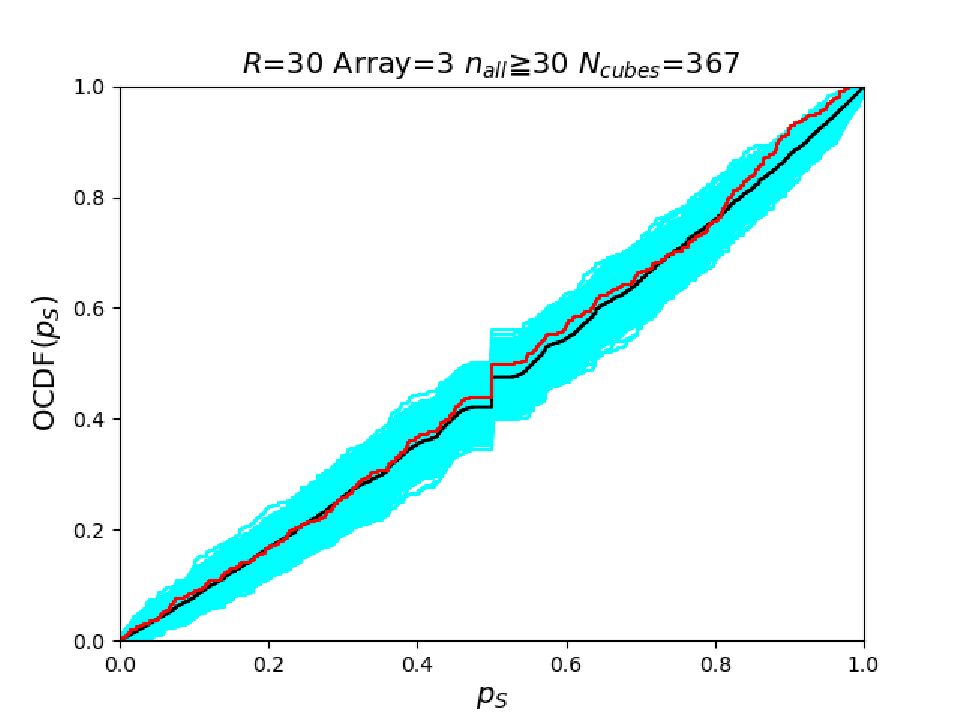}
\caption{
 (Upper panel) Observed cumulative distribution function $OCDF$ for Cube Set S1. In this case, the observed $OCDF$ extends beyond the range covered by 1000 Monte Carlo simulations. The discontinuity at $p_{\rm S}=0.5$ in the theoretical distribution is less pronounced due to the large number of galaxies $n_{\rm all} \ge 300$ in the cubes selected for this set.
 (Lower panel) $OCDF$s for Cube Set S127 (see Table \ref{tab: KS statistic}).  The observed distribution (red), the theoretical binomial expectation (black), and the results from 1000 Monte Carlo simulations (cyan) are shown. The observed $OCDF$ lies well within the range spanned by the Monte Carlo simulations.}
\label{fig: p-f(p)}
\end{figure}

We computed the observed values of $p_{\rm S}$ and $p$ for each of the 46,247 searched cubes. We also generated 1000 Monte Carlo realizations by randomly assigning S/Z parity to each of the 49,494 spiral galaxies, and calculated the corresponding $p_{\rm S}$ and $p$ for each realization. This ensemble is used in the next sections to assess the consistency of the observed S/Z spin distribution with a random model.

\subsection{KS Test of Observed Cube Sets}

For a set of $N^{k}_{\rm cubes}$ independent searched cubes ($i = 1, \ldots, N^{k}_{\rm cubes}$), each containing $n_{\rm all} \ge n_{\rm lim}$ annotated spirals, 
we construct the $OCDF(p_{\rm S})$, in the range $2^{-\max(n_{\rm all}^{i})} \le p_{\rm S} \le 1$, by summing, over all cubes, the maximum $p^{i}_{{\rm S},k}$ satisfying $p^{i}_{{\rm S},k} \le p_{\rm S}$, and normalizing by $N^{k}_{\rm cubes}$, 
\begin{eqnarray}
OCDF(p_{\rm S})=\frac{1}{N^{k}_{\rm cubes}} \sum_{i=1}^{N^{k}_{\rm cubes}}\max(p^{i}_{\rm S,k} | p^{i}_{\rm S,k} \leq p_{S}). 
\end{eqnarray}

Figure \ref{fig: OCDF} illustrates how this binomial $OCDF$ for a set of cubes deviates from the diagonal line that corresponds to the $OCDF$ of a normal distribution.

To quantify the departure of the observed $OCDF$ for a cube set $k (1 \le k \le 256)$, we construct a fiducial theoretical cube set in which each cube with $n^{i}_{\rm all}$ galaxies follows a purely binomial distribution, and compute the corresponding theoretical $OCDF$. We then compare the observed and theoretical $OCDF$s, and evaluate the Kolmogorov–Smirnov (KS) statistic, $D^{k}_{\rm obs}$, as
\begin{equation}
D^{k}_{\rm obs} = \max \{ |OCDF^{k}_{\rm obs}(p_{\rm S})- OCDF^{k}_{\rm theo}(p_{\rm S})| \}.
\end{equation}

Similarly, we calculate KS statistic $D^{k}_{j}$ ($j=1, \ldots, 1000$) for the 1000 Monte Carlo realizations, forming a matrix $D_{j}^{k}$($j=1, \ldots, 1000$; $k=1, \ldots, 256$). We define a significance measure
\begin{eqnarray}
\sigma^{k}_{D} = \frac{D^{k}_{\rm obs} - \langle D^{k}\rangle}{D^{k}_{\rm rms}},
\end{eqnarray}
with the average $\langle D^{k}\rangle$ and the dispersion $D^{k}_{\rm rms}$ computed from the 1000 random realizations. The 256 cube sets are then sorted by decreasing $\sigma_{D}^{k}$ and relabeled accordingly as Set1 to Set256.

Table \ref{tab: KS statistic} shows a part of the summary data. Among the 256 sets, we find three outstanding cube sets where $D^{k}_{obs} > D^{k}_{\rm max}$.

The upper panel of Figure \ref{fig: p-f(p)} shows the observed $OCDF$ (red), theoretical $OCDF$(black), and 1000 Monte Carlo $OCDF$s(cyan) for the cube set Set1, where the observed curve deviates from the distribution of 1000 simulated curves. The red curve above the black curve corresponds to the dominance of Z-spiral over S-spiral in number,

The lower panel of Figure \ref{fig: p-f(p)} shows, as an illustrative example, the $OCDF$s, for a typical cube S127 with median value of $D^{k}_{j}$ among the 256 sets. The central jump at $p_{\rm S}=0.5$ in the OCDFs corresponds to cubes for which $n_{\rm S}=n_{\rm Z}-1$. In general, $p_{\rm S}$ takes various values in cubes with different $n_{\rm all}$, while if $n_{\rm all}$ is odd and $n_{\rm S} = n_{\rm Z}-1$, $p_{\rm S} = 0.5$ for all cubes and this makes a concentration of $p_{\rm S}$ there.

\begin{table}[] 
\centering
\caption{The KS statistics for Cube Sets sorted in order of the significance measure $\sigma_{D}$.}\vspace{-7pt}
\begin{tabular}{cccccccc}\hline
SetID&$R$&Array&$n_{\rm lim}$&$N_{\rm cubes}$&$D_{\rm obs}$&$D_{\rm max}$&$\sigma_{D}$\\ \hline
Set1&100&1&300&25&0.423&0.361&5.29\\ \hline
Set2&37.5&7&30&327&0.110&0.103&4.64\\ \hline
Set3&60&6&100&101&0.196&0.179&4.46\\ \hline
Set4&100&4&30&25&0.369&0.405&4.05\\ \hline
...&...&...&...&...&...&...\\ \hline
Set127&30&3&30&367&0.051&0.097&0.57\\ \hline
...&...&...&...&...&...&...\\ \hline
\end{tabular}
\label{tab: KS statistic}
\end{table}

\subsection{Statistical Significance of Outstanding Cube Sets as a Whole}
The probability of finding 3 outstanding cube sets among 256 sets, where the observed KS statistic $D_{\rm obs}$ exceeds $D_{\rm max}$ obtained from 1000 Monte Carlo simulations, is given by $1-p \bigl(Bi(256,0.001) \le 2\bigr) = 0.0023$, if the 256 sets are mutually independent. However, because the 256 cube sets are not independent, we evaluated the statistical significance instead using the 1000 random simulations themselves, treating each of the 256 sets in a consistent manner.

We select one realization from the 1000 Monte Carlo simulations, $j=J$, and treat $D^{k}_{j=J}$ as a fiducial observed KS value. The remaining 999 values, $D^{k}_{j \neq J}$, are then used as the reference distribution to evaluate this fiducial observation. Repeating this procedure for every realization $j$, we count $\mathscr M_{\rm run}$, the number of Monte Carlo runs in which $\mathscr N_{\rm out}$  cubes sets satisfy $D^{k}_{J} > \max\left(D^{k}_{j \neq J}\right)$. Table \ref{tab: KS calib} presents $\mathscr M_{\rm run}$ among the 1000 realizations as a function of the number of outstanding cube sets $\mathscr N_{\rm out}$.

The 24 runs which yield $\mathscr N_{\rm out} \ge 3$ in 1000 correspond to 2.4\%.  This indicates that the observed spin distribution of galaxies is consistent with a random binomial distribution with a risk at the 2.4 \% level.

In addition to the statistical tests described above, we conducted separate simulations where the annotation probability of spiral handedness was artificially tuned to match the observed slight excess of Z-spirals, $24,891/46,247=0.5029$. While this adjustment alleviates the discrepancy in individual cases (e.g., cube set Set1), it leads to inconsistencies across the remaining cube sets, showing an excess of Z-spirals.

If the 2.4\% possibility is taken at face value, one might speculate that the deviation could be linked to a large-scale phenomenon -- such as baryon acoustic oscillation \citep[e.g.,][]{Eisenstein2005}, or parity violation associated with dark matter and/or dark energy \citep[e.g.,][]{Komatsu2022}.  However, we currently have no compelling physical interpretation for this potential small anomaly.

\begin{table}[] 
\centering
\caption{The number of random realizations $\mathscr M_{\rm run}$ out of 1000 that produce $\mathscr N_{\rm out}$ outstanding cube sets.}
\begin{tabular}{ccccccccccc} \hline
$\mathscr N_{out}$&0&1&2&3&4&5&8&11&24&$(\mathscr N_{\rm out} \ge 3)$\\ \hline
$\mathscr M_{run}$&865&97&24&13&5&2&2&1&1&24\\ \hline
\end{tabular}
\label{tab: KS calib}
\end{table}

\begin{figure}[h] 
\centering
\includegraphics[scale=0.45]{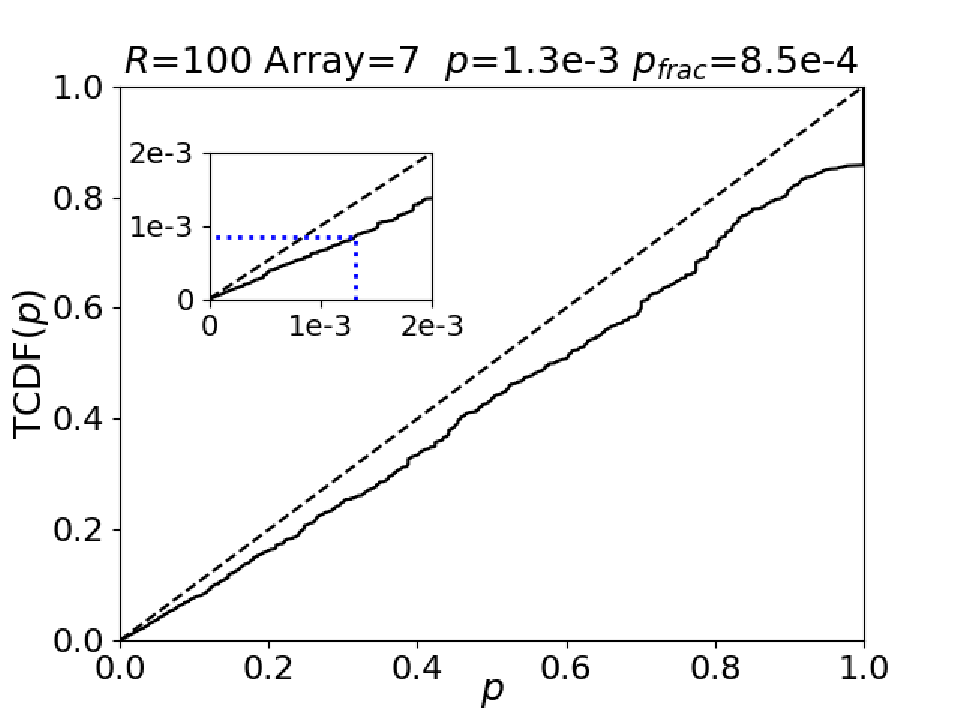}
\caption{The solid curve shows the theoretical two-sided binomial cumulative distribution function $TCDF(p)$ for a cube set with $R=100$ in Array 7. The probability of obtaining an asymmetry equal to or larger than that observed for this set is given by $p_{\rm frac} = TCDF(p)$.  The dashed line represents the corresponding $TCDF$ for the normal distribution. In the binomial case, the curve deviates from the normal distribution due to its discrete and finite nature.}
\label{fig: p_frac}
\end{figure}

\begin{table*}[] 
\centering
\caption{Data for the bizarrest cube; details are provided in the text.}
\begin{tabular}{ccccccccccccc} \hline
ID&SGXC&SGYC&SGZC&$R$&$n_{\rm S}$&$n_{\rm Z}$&$n_{\rm all}$&$p$&$N^{k}_{\rm cubes}$&$p_{\rm frac}$&$b$\\ \hline
48016&262.5&-788&675&37.5&12&44&56&2.09e-05&601&8.40e-06&0.0050\\ \hline
\end{tabular}
\label{tab: BizarreCubes}
\end{table*}

\subsection{``Bizarre'' Cubes}

Next, we investigate whether any cube exhibits deviations from a random distribution that are not captured by the KS statistics.
We define the two-sided cumulative distribution function $TCDF(p)$ in an analogous way, by replacing $p_{\rm S}$ to $p$ in equation (5) for 80 cube sets with $N_{cubes} \ge 11$.
Thus defined $TCDF(p)$ increases in a stair-like way, from 0 to 1, as shown by the black curve in  Figure \ref{fig: p_frac}, for a cube set with $N_{\rm cubes} = 27$ and $R=100$ in Array 7.

We calculate the expected fraction of cubes in the set that has $p$-value equal to or smaller than the $p$-value of a certain cube. We define the fraction as $p_{\rm frac}(p)$, which is equivalent to $TCDF(p)$, the normalized two-sided cumulative distribution function of $p$. If the distribution of $p$ is continuous, as is the case for the Gaussian distribution, $TCDF(p) = p$. In this study, however, the $TCDF(p)$ is discrete and $p_{\rm frac}(p^{i}) = TCDF(p^{i}) \le p^{i}$.

We then define the ``bizarreness parameter'' $b^{i}$ as
\begin{equation}
b^{i}=p_{\rm frac}^{i} N^{k}_{\rm cubes},
\end{equation}
which corresponds to the expected number of cubes that have $p$-value equal to or smaller than $p$. Therefore, a value $b^{i}< 1$ implies that such an outcome is nominally unlikely to occur. For $b \ll 1$, the probability that a cube with $b$ exists is $1- \exp (-b) \sim b$.

\begin{figure}[] 
\centering
\includegraphics[scale=0.23]{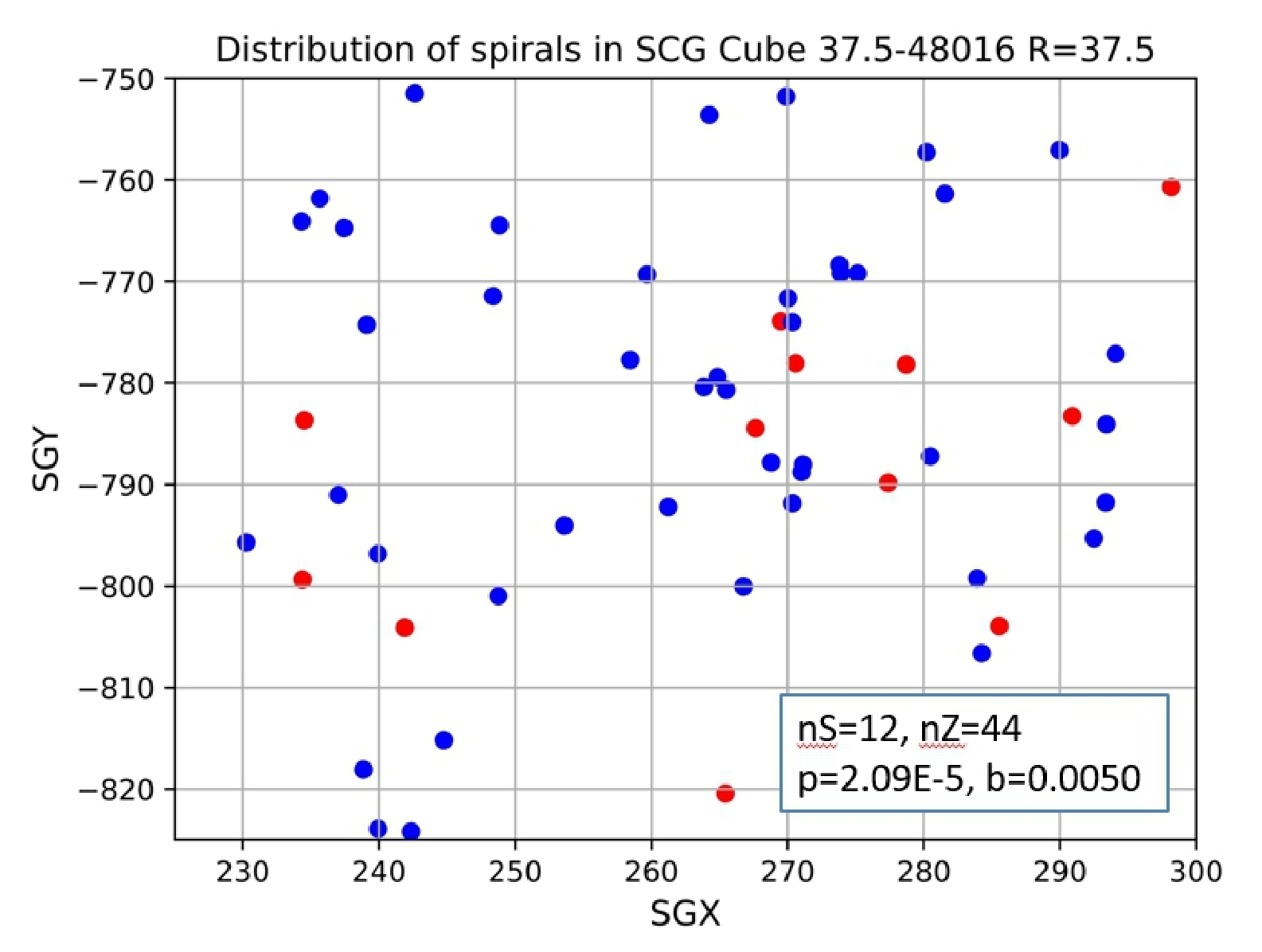}
\caption{Spatial distribution of S-spirals (red) and Z-spirals (blue) galaxies in the SGX-SGY plane for  Cube37.5-48016.}
\label{fig: BizarreCube}
\end{figure}

\begin{table}[] 
\centering
\caption{Number of bizarre cubes in the observed data, the theoretical independent binomial distribution. The bottom row shows the mean and the standard deviation of the number of bizarre cubes, based on 1000 random Monte Carlo realizations.} \vspace{-7pt}
\begin{tabular}{cccc} \hline
Sample&$b < 1$&$b < 0.1$&$b < 0.01$\\ \hline
Observed&72&9&1\\ \hline
Theoretical&80&8&0.8\\ \hline
Monte Carlo Runs&78.1$\pm$16.7&7.3$\pm$3.9&0.7$\pm$1.1\\ \hline
\end{tabular}
\label{tab: Cube-p-R}
\end{table}

Table \ref{tab: BizarreCubes} lists the data for the ``bizarrest'' cube. Figure \ref{fig: BizarreCube} presents the SGX-SGY projected spatial distribution of S-spirals (red) and Z-spirals (blue) for this cube.

Among the 46,247 searched cubes, we identify 72, 9, and 1 cubes, with $b<1, \ 0.1$, and  0.01, respectively. Table \ref{tab: Cube-p-R} summarizes the corresponding numbers expected from the theoretical binomial distribution and from 1000 random Monte Carlo simulations. The observed counts are consistent with both the theoretical expectations and the simulation results, indicating that even the bizarrest cube does not contradict with a random distribution.

\begin{table*}[t] 
\centering
\caption{Data for the bizarrest hemicube. The section plane used to divide the cube into two hemicubes is given in the last column.}
\begin{tabular}{ccccccccccccc} \hline
ID&SGX&SGY&SGZ&$R$&$n_{\rm S1}$&$n_{\rm Z1}$&$n_{\rm S2}$&$n_{\rm Z2}$&$p^{\rm hemi}$&$b^{\rm hemi}$&Section plane&\\ \hline
12231&360&-900&480&60&13&5&48&98&6.94e-06&0.018&SGX=360&\\ \hline
\end{tabular}
\label{tab: BizarreHemiCubes}
\end{table*}

\subsection{HemiCube analysis}
Even when a cube contains an equal number of S-spirals and Z-spirals ($n_{\rm S}=n_{\rm Z}$),  the two populations may still be spatially separated, forming a dipole-like distribution.
To investigate the presence of any significant hemisphere asymmetry, we performed an additional analysis by dividing each cube into two hemicubes using nine different partition planes.
These planes are defined by $SGX=0$, $SXY=0$, $SGZ=0$, $SGX=SGY$, $SGX=-SGY$, $SGY=SGZ$, $SGY=-SGZ$, $SGZ=SGX$, and $SGZ=-SGX$. We counted the number of S-spirals and Z-spirals, $n_{\rm S1}, n_{\rm S2}, n_{\rm Z1}$, and $n_{\rm Z2}$, in each of the hemicubes 1 and 2.

Instead of $n_{\rm S}$ and $n_{\rm Z}$ used in Section 3.3 to define the two-sided tail probability $p$, we adopt $n_{\rm S1}+n_{\rm Z2}$ and $n_{\rm S2}+n_{\rm Z1}$ to define the two-sided hemicube probability $p^{\rm hemi}$ for divided hemicubes.

\begin{figure}[h] 
\centering
\includegraphics[scale=0.2]{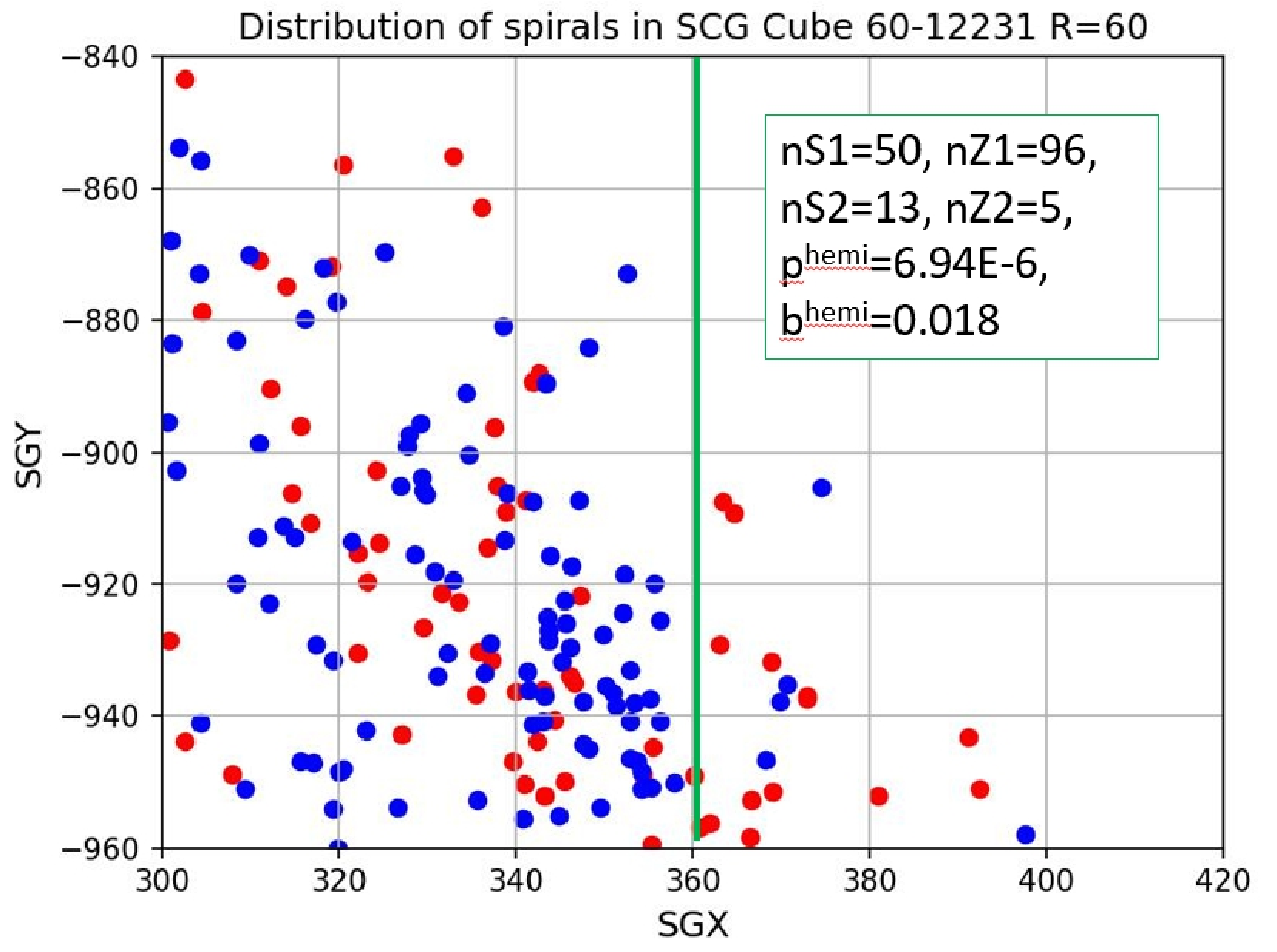}
\caption{Distribution of S-spirals (red) and Z-spirals (blue) in the bizarrest hemicube. The hemicube on the left of the green section plane contains a higher number of Z-spirals, while the other hemicube on the right contains a higher number of  S-spirals.}
\label{fig: BizarreHemiCube}
\end{figure}

Among the 46,247 cubes, each examined with nine sectioning planes, Table \ref{tab: BizarreHemiCubes} summarizes the bizarrest hemicube asymmetries, quantified by $b^{\rm hemi}=9 p^{\rm hemi}N^{k}_{\rm cubes}$.   As an example, Figure \ref{fig: BizarreHemiCube} shows a pronounced number asymmetry of 111 versus 53 galaxies across the SGX = 360 Mpc dividing plane for Cube-60-12231. The corresponding binomial probability is $p^{\rm hemi} =6.94 \times 10^{-6}$, yielding $b^{\rm hemi} = 0.018$. As in the case of bizarre cubes discussed in the previous section, the statistics for bizarre hemicubes indicate that the overall occurrence rate is consistent with the theoretical binomial distribution.

Lastly, we comment on the connection between the present hemicube-asymmetry analysis and a possible dipole asymmetry in the three-dimensional spin-vector distribution. Because the S/Z parity encodes only the line-of-sight component of a galaxy's spin vector, the present method is sensitive exclusively to a dipole-like asymmetry with radial direction. Conversely, any dipole spin asymmetry with an axis orthogonal to the radial axis would remain undetectable by this analysis.

\section{Summary and Discussion} \label{sec:summary}

To test for possible departures from randomness in galaxy spin vectors, we analyzed the distribution of 49,494 spiral galaxies with spectroscopic redshifts and S/Z spiral annotations within the footprint of the Subaru HSC WIDE survey. 
We confirmed that the distribution of the spin is mostly consistent with the binomial distribution as a whole.

We identified a small number of ``bizarre'' cubes in which the probability of obtaining the observed asymmetry between S- and Z-spiral galaxies is lower than that expected from the theoretical binomial distribution. 
However, the spin-vector distribution of 49,494 spiral galaxies examined in the present HSC-WIDE dataset -- extending to distances of approximately 2 Gpc -- indicates that the observed asymmetry between $n_{\rm S}$ and $n_{\rm Z}$ in galaxies samples drawn from fiducial search cubes of size $20 \text{-}200$ Mpc is statistically consistent with a random binomial distribution at 2.4 \% level.

This result is compatible with the standard $\Lambda$CDM model, in which vorticity-free perturbations grow into large-scale structure through gravitational instability, and with tidal torquing theory, which predicts that galaxy spins are distributed randomly. Although the present study does not directly test spatial correlations in spin distribution, any systematic bias in the number asymmetry of S/Z spirals, if present, must have been substantially suppressed or erased during the subsequent nonlinear dynamical processes of structure formation that shaped the present-day galaxy distribution.

\begin{acknowledgments}

The authors thank Kenichi Tadaki for providing the CNN data from \citet{Tadaki2020} and Toshiki Kurita for valuable information on studies of intrinsic shape. Rinon Kageyama contributed to preparing the SZ data for the verification sample. Critical readings and valuable suggestions of the original and the revised texts by an anonymous referee are highly appreciated. This research was supported by JSPS KAKENHI Grant Number 22K03679 and was carried out, in part, using the Multi-wavelength Data Analysis System operated by the Astronomy Data Center (ADC), National Astronomical Observatory of Japan.
This research also made use of NASA's Astrophysics Data System Bibliographic Services and the VizieR catalogue access tool, CDS, Strasbourg, France (DOI: 10.26093/cds/vizier).
\rm
\end{acknowledgments}

\appendix
\section{Spin catalog of Galaxies}
\label{sec:catalog}

We cross-matched all the 76,636 spirals that have spin information
as S-wise or Z-wise with spectroscopic data.
The following data catalogs were obtained from VizieR:
The 2dF Galaxy Redshift Survey 
\citep[2dF;][]{Colless2001},
6dF galaxy survey final redshift release
\citep[6dF;][]{Jones2004,Jones2009},
Galaxy And Mass Assembly(GAMA) Data Release 3
\citep[GAMADR3;][]{Baldry2018},
The VIMOS VLT Deep Survey(VVDS) final data release
\citep[VVDS;][]{LeFevre2013},
The WiggleZ Dark Energy Survey
\citep[WiggleZ;][]{Drinkwater2018},
and
The VIMOS Public Extragalactic Redshift Survey (VIPERS)
Public Data Release 2
\citep[VIPERS;][]{Scodeggio2018}.
The PRIsm MUlti-Object Survey (PRIMUS) Data Release 1
\citep[PRIMUS;][]{Coil2011,Cool2013}
was obtained from 
\url{https://primus.ucsd.edu/version1.html}.
Dark Energy Spectroscopic Instrument (DESI) Data Release 1
\citep[DESIDR1;][]{DESI2025}
was obtained from
\url{https://data.desi.lbl.gov/public/dr1/}.
Data of Sloan Digital Sky Survey (SDSS) Data Release 17
\citep[SDSSDR17;][]{Abdurrouf2022}
were obtained via SciServer 
(\url{https://skyserver.sdss.org/casjobs/}).
We also used our new results from Subaru Prime Focus Spectrograph observation (PFS-S25A).

If an object has multiple redshifts, we checked whether they are consistent. If they are consistent, we adopted redshifts from DESI, SDSS, GAMA, and others in this priority order.
If an inconsistency is found, we visually examined the spectra of DESI or SDSS, and judged which one is relevant.
For some objects in DESI and SDSS, we noticed that the catalog redshift does not well fit with the spectroscopic data.
We manually corrected the redshift to match the spectra for such cases. They are indicated as CORR-DESIDR1 and CORR-SDSSDR17.
We also removed $z<0.01$ objects, since their supergalactic coordinate
would not be determined relevantly from the redshift.

Table \ref{tab:catalog} gives the RA, Dec, and i-band magnitude, adopted redshift, and the redshift source of 49,494 galaxies.
The table also gives the calculated supergalactic coordinates.

\begin{deluxetable}{cccc}
\caption{Spin Catalog}
\label{tab:catalog}
\tablehead{
   \colhead{Row number} & 
   \colhead{Label} & 
   \colhead{Units} & 
   \colhead{Description}
}
\startdata
1& ID & --- & HSC-SSP PDR2 object ID\\
2& RA & deg & Right Ascension in decimal degrees (J2000) \\
3& Dec & deg & Declination in decimal degrees (J2000) \\
4& i58apmag & mag & i-band aperture magnitude in 5.8 arcsec\\
5& spin & --- & 1:S-wise 2:Z-wise\\
6& specsrc & --- & source of specz\tablenotemark{*} \\
7& specname & --- & name of the specrtoscopic target \\
8& specra & --- & right ascension of spectroscopic target in decimal degrees (J2000) \\
9& specdec & --- & declination of  spectroscopic target in decimal degrees (J2000) \\
10& specz & --- & spectroscopic redshift \\
11& SGX & Mpc & supergalactic coordinate X \\
12& SGY & Mpc & supergalactic coordinate Y \\
13& SGZ & Mpc & supergalactic coordinate Z \\
\enddata
\tablenotetext{*}{ 
One of 
2dF, 
6dF, 
GAMADR3, 
VVDS, 
WiggleZ, 
VIPERS, 
PRIMUS, 
DESIDR1, 
SDSSDR17, 
PFS-S25A, 
CORR-DESIDR1, 
or CORR-SDSSDR17 
is used. 
The details are given in the text.}
\tablecomments{Table \ref{tab:catalog} is published in its entirety in the
machine-readable format. A portion is shown here for guidance regarding its form and content.}
\end{deluxetable}

\bibliographystyle{aasjournalv7}

\begin{thebibliography}{}
\bibitem[A. H. Abdullah\& P. Kroupa(2014)]{Abdullah2014} Abdullah, A. H. and Kroupa, P. 2014, ASP Conf. 486, 137

\bibitem[Abdurro'uf et al.(2022)]{Abdurrouf2022}
Abdurro'uf \etal 2022, \apjs, 259, 35 

\bibitem[H. Aihara et al.(2019)]{Aihara2019} Aihara, H., AlSayyad,Y., Ando, M. \etal 2019, \pasj, 71, 114

\bibitem[I. K. Baldry et al.(2018)]{Baldry2018}
Baldry I. K. \etal 2018, \mnras, 474, 3875 
\bibitem[J. Bardeen(1980)]{Bardeen1980} Bardeen, J. M. 1980, Physics Review  D, 22, 1882.
\bibitem[A. J. Christopherson(2011)]{Christopherson2011} Christopherson, A. J. 2011, arXive:1106.0446v1
\bibitem[A. L. Coil et al.(2011)]{Coil2011}
Coil, A.L. \etal 2011, \apj, 741, 8 

\bibitem[M. Colless et al.(2001)]{Colless2001} Colless, M., \etal 2001, \mnras, 328, 1039

\bibitem[R. J. Cool et al.(2013)]{Cool2013}
Cool, R.J. \etal 2013, \apj, 767, 118 
\bibitem[DESI Collaboration et al.(2025)]{DESI2025}
DESI Collaboration \etal 2025, arxiv:2503.14745
\bibitem[M. J. Drinkwater et al.(2018)]{Drinkwater2018}
Drinkwater, M. J. \etal 2018, \mnras, 474, 4151
\bibitem[D. J. Eisenstein et al.(2005)]{Eisenstein2005}
Eisenstein, D. J. \etal 2005, \apj, 633, 560 
\bibitem[W. B. Hayes et al.(2017)]{Hayes2017} Hayes, W. B., Davis, D., and Silva, P. 2017, \mnras, 466, 3928

\bibitem[C. M. Hirata \& U. Seijak(2004)]{Hirata2004} Hirata, C. M. \& Seljak, U. 2004, \prd, 70, 063526

\bibitem[C. M. Hirata et al.(2007)]{Hirata2007} Hirata, C. M., Mandelbaum, R., Ishak, M. \etal 2007, \mnras, 381, 1197
\bibitem[M. Iye et al.(2019)]{Iye2019} Iye, M., Tadaki, K., and Fukumoto, H. 2019, \apj, 886, 133
\bibitem[H. Johnston et al.(2019)]{Johnston2019} Johnston, H., Geprgiou, C., Joachimi, B. \etal 2019, \aa, 624, A30

\bibitem[D. H. Jones et al.(2009)]{Jones2009}
Jones, D. H. \etal 2009, \mnras, 399, 683 
\bibitem[D. H. Jones et al.(2004)]{Jones2004}
Jones, D. H. \etal 2004, \mnras, 355, 747 
\bibitem[E. Komatsu(2022)]{Komatsu2022}
Komatsu, E. 2022, Nature Reviews Physics, 4, 452 
\bibitem[O. Le Fevre et al.(2013)]{LeFevre2013}
Le Fevre, O. \etal 2013, \aap, 559, A14 

\bibitem[C. Li et al.(2013)]{Li2013} Li, C., Jing, Y.P., Faltenbacher, A., and Wang, J. 2013, \apjl, 770, L12 
\bibitem[N. I. Libeskind et al(2013)]{Libeskind2013} Libeskind, N. I., Hoffman, Y., Steinmetz, M. \& Gottl''{o}ber, S. 2013 \apj, 766, L15.
\bibitem[C. Lintott et al.(2011)]{Lintott2011} Lintott, C., Schawinski, K., Bamford, S. \etal 2011, \mnras, 410, 166
\bibitem[P. Lopez et al.(2021)]{Lopez2021} Lopez, P., Cautun, M., Paz, D. \etal. 2021, \mnras, 502, 5528

\bibitem[R. Mandelbaum et al.(2011)]{Mandelbaum2011} Mandelbaum, R., Blake, C, Bridle, S. \etal 2011, \mnras, 410, 844
\bibitem[P. Motloch et al.(2022)]{Motloch2022} Motloch, P., Pen, U.-L., and Yu, H.-R., 2022, Phys. Rev.. D, 105, 083504 
\bibitem[P. Pajowska et al.(2019)]{Pajowska2019} Pajowska, P., Godfowski, W., Zhu, Z. \etal 2019, J. Cosmology Astroparticle Physics, 2, 5. 
\bibitem[D. Patel \& H. Desmond(2024)]{Patel2024} Patel, D. \& Desmond, H. 2024, \mnras, 534, 1553
\bibitem[P. J. E. Peebles(1969)]{Peebles1969} Peebles, P. J. E. 1969, \apj, 155, 393
\bibitem[Planck Collaboration et al.(2020)]{Aghanim2020} Planck Collaboration, 
Aghanim, N., Akrami, Y., Ashdown, M. \etal 2020, \aap, 641, A6

\bibitem[M. Scodeggio et al.(2018)]{Scodeggio2018}
Scodeggio, M. \etal 2018, \aap, 609, A84 

\bibitem[K. Tadaki et al.(2020)]{Tadaki2020} Tadaki, K., Iye, M., Fukumoto, H., Hayashi, M., Rusu, C. E., Shimakawa, R. and Tosaki T. 2020, \mnras, 496, 4276
\bibitem[I. Trujillo et al.(2006)]{Trujillo2006} Trujillo, I., Carretero, C., and Patiri, S.G. 2006, \apj, 640, L111 
\bibitem[J. Varela et al.(2012)]{Varela2012} Varela, J., Betancort-Rijo, J.,Trujillo, I. and Ricciardelli, E. 2012, \apj, 744, 82 
\bibitem[S. White(1984)]{White1984} White, S., 1984, \apj, 286, 38
\bibitem[K. W. Willett et al.(2013)]{Willett2013} Willett K. W., Lintott, C.J., Bamford, S.P. \etal 2013, \mnras, 435, 2835 
\bibitem[M. Yagi et al.(2025)]{Yagi2025} Yagi, M., Iye, M. and Fukumoto, H., 2025, \aj, 170, 90.
\bibitem[D. G. York et al.(2000)]{York2000} York, D.G. \etal 2000, \apj, 120, 1579
\bibitem[Y. Zhang et al.(2013)]{Zhang2013} Zhang, Y., Yang, X., Wang, H. \etal 2013, \apj, 779, 160 
\bibitem[Y. B. Zeldovich(1970)]{Zeldovich1970} Zeldovich, Y. B., 1970, \aap, 5, 84
\end{thebibliography}

\end{document}